\documentclass[runningheads]{llncs}
\usepackage[T1]{fontenc}
\usepackage{graphicx} 

\usepackage[acronym]{glossaries}
\glsdisablehyper
\newacronym{ot}{OT}{Operational Technology}
\newacronym{vpn}{VPN}{Virtual Private Network}
\newacronym{ics}{ICS}{Industrial Control System}
\newacronym{rdp}{RDP}{Remote Desktop Protocol}
\newacronym{plc}{PLC}{Programmable Logic Controller}
\newacronym{rtu}{RTU}{Remote Terminal Unit}
\newacronym{apt}{APT}{advanced persistent threat}
\newacronym{scada}{SCADA}{Supervisory Control and Data Acquisition}
\newacronym{hmi}{HMI}{Human Machine Interface}
\newacronym{asn}{ASN}{Autonomous System Number}
\newacronym{api}{API}{Application Programming Interface}
\newacronym{ocr}{OCR}{Optical Character Recognition}
\newacronym{cve}{CVE}{Common Vulnerabilities and Exposure}
\newacronym{ews}{EWS}{Engineering Workstation}
\newacronym{it}{IT}{Information Technology}
\newacronym{io}{I/O}{Input/Output}
\newacronym{iot}{IoT}{Internet of Things}
\newacronym{mppt}{MPTT}{Maximum Power Point Tracking}
\newacronym{isp}{ISP}{Internet Service Provider}
\newacronym{cip}{CIP}{Common Industrial Protocol}
\newacronym{dnp3}{DNP3}{Distributed Network Protocol v3}
\newacronym{tls}{TLS}{Transport Layer Security}
\newacronym{gui}{GUI}{Graphical User Interface}

\usepackage{courier} 
\newcommand{\queryfmt}[1]{\texttt{#1}}

\begin{document}

\title{Analysis of Publicly Accessible Operational Technology and Associated Risks}
\titlerunning{Analysis of Publicly Accessible OT and Associated Risks}
\author{Matthew Rodda\inst{1}\orcidID{0000-1111-2222-3333} \and
Vasilios Mavroudis\inst{2}\orcidID{1111-2222-3333-4444}}
\authorrunning{M. Rodda et al.}
%
\institute{Defence Science and Technology Group, Australia \\
\email{matt.rodda@defence.gov.au} \and
The Alan Turing Institute, United Kingdom}

\maketitle
\begin{abstract}
\gls{ot} is an integral component of critical national infrastructure, enabling automation and control in industries such as energy, manufacturing, and transportation.
However, OT networks, systems, and devices have been designed and deployed prioritising functionality rather than security.
This leads to inherent vulnerabilities in many deployed systems when operational misconfigurations expose them to the internet. 
This report provides an up-to-date overview of the OT threat landscape exposed to the public internet and studies the affected protocols, vendors, software, and the geographic distribution of systems. 
Our findings reveal nearly 70,000 exposed OT devices globally, with significant concentrations in North America and Europe. Analysis of prevalent protocols (e.g., ModbusTCP, EtherNet/IP, S7) shows that many devices expose detailed identifying information, including outdated firmware versions with known critical vulnerabilities that remain unpatched for years after disclosure.
Furthermore, we demonstrate how automated analysis of screenshots can uncover exposed graphical interfaces of \glspl{hmi} and \gls{scada} systems, highlighting diverse pathways for potential unauthorized access and underscoring the risks to industrial processes and critical infrastructure.

\end{abstract}

\glsresetall
\section{Introduction}

\gls{ot} forms the backbone of national infrastructure, automating physical processes across critical sectors including utilities, manufacturing and transportation~\cite{stouffer2023guide}.
It is essential that such systems operate reliably and safely to ensure public well-being and economic stability.
However, \gls{ot} systems have historically been designed with a primary focus on reliability and safety, rather than cybersecurity~\cite{jillepalli2017security}.
As a result, \gls{ot} networks often lack the fundamental security mechanisms that would be expected of an equivalently critical \gls{it} system.
Further, malicious actors are increasingly targeting \gls{ot} systems, including sophisticated state-sponsored groups conducting espionage or attempting to disrupt essential services~\cite{dragos2025,dragosgroups2025}.
Landmark attacks such as Stuxnet, Industroyer/Industroyer2, and INCONTROLLER illustrate the unique potential for physical damage incurred by \gls{ot} attacks~\cite{kushner2013real,kozak2023industroyer,boakye2023securing}.
To mitigate against such attacks, it is essential that \gls{ot} network administrators regularly patch their devices and implement strict access-control policies.

Regular patching of \gls{ot} devices ensures that known vulnerabilities cannot be utilized to compromise the device.
This greatly increases the time and skill required of a potential attacker to compromise the device, as they cannot apply a known attack but instead must develop a novel (zero-day) attack.
- in this study we demonstrate that OT asset owners often do not keep devices up to date
- we conjecture that this may be due to the downtime imposed by patches and perceived risk of interrupting operations




To mitigate these risks, organisations often rely upon cybersecurity policies focussing on perimeter security, such as network segmentation, firewall protections, and traffic monitoring.
This ``walled garden'' approach aims to prevent threats from reaching critical systems rather than securing the systems themselves.
Such a strategy is pragmatic as modifications to the \gls{ot} environment that would enhance their security may be prohibitively expensive.
However, a walled garden approach introduces a single point of failure, where the perimeter's compromise subsequently reveals defenceless \gls{ot} devices.

Despite awareness of these risks, the actual extent and characteristics of OT/ICS assets exposed to the public internet represents a critical knowledge gap.
This report provides a contemporary, data-driven analysis of publicly accessible OT systems.
Leveraging internet-wide scanning data, we investigate the volume, geographic distribution, protocols, vendors, and firmware of exposed devices, with a specific focus on \glspl{ics}.
By identifying and analysing the exposed attack surface, we seek to quantify the tangible risks posed by internet-facing OT.
This work complements previous studies with up-to-date information as the device distribution changes and threats continually evolve~\cite{mirian2016internet}.

\section{Background and Related Works}
\begin{figure}[!htbp]
\centering
\includegraphics[width=.75\linewidth,trim={1cm 8cm 1cm 1cm},clip]{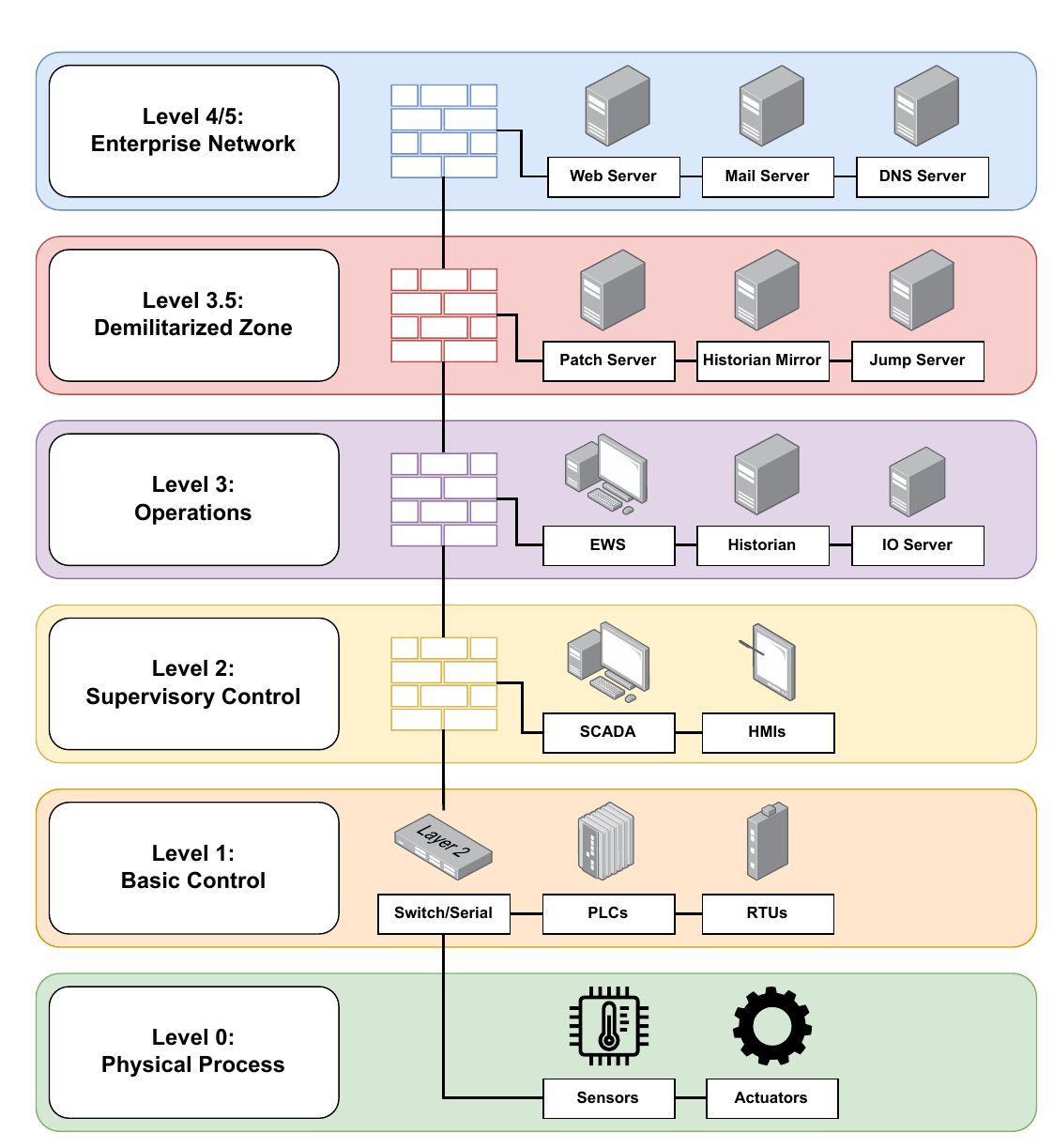}
\caption{The Purdue Model for Control Hierarchy, illustrating segmentation within a typical OT/ICS environment~\cite{purdue-model}}
\label{fig:purdue-model}
\end{figure}

\Gls{ot} encompasses the wide range of hardware and software systems used to monitor and control physical processes, devices, and infrastructure.
\glspl{ics} comprise a large and critical subset of \gls{ot}, specifically referring to systems used for managing industrial processes, often incorporating \glspl{plc}, \glspl{hmi} and \gls{scada} systems.
A common framework used to describe the hierarchy and data flow within \gls{ot} environments, particularly in industrial settings, is the Purdue Model~\cite{purdue-model}.
As shown in Figure~\ref{fig:purdue-model}, this model segments systems into logical levels based on their function, from the physical process itself up to the enterprise network.
Level 0 of the Purdue Model represents the physical process and the devices that directly interact with it, such as sensors and actuators.

Level 1 consists of the devices responsible for direct automation and control of Level 0 equipment, through specialised controllers such as \glspl{plc} and \glspl{rtu}.
These devices read sensor inputs and execute control logic to manipulate actuators, often operating under strict real-time constraints.
They differ significantly from typical IT devices due to requirements for: 1) Robustness to harsh industrial conditions, 2) Reliability during continuous, long-term operation, 3) Real-time performance with low and predictable latency for process control, and 4) Specialised \gls{io} to interface with a wide variety of industrial devices as well as typical IT network infrastructure.
These unique design constraints mean they cannot simply be replaced by conventional IT hardware, despite the latter offering superior native security features.

Level 2 of the Purdue Model involves supervisory control functions.
Systems at this level aggregate data from Level 1 controllers, provide visualisation for human operators, and allow for centralised monitoring and control.
Common examples include \glspl{hmi}, which offer graphical interfaces for operators to interact with the process, and \gls{scada} systems, used for monitoring system state of large-scale processes, especially within geographically dispersed applications such as pipelines or power distribution.
Level 3 includes systems supporting site-wide manufacturing or operational management (e.g., historians, optimisation systems, scheduling systems, engineering workstations).
These systems bridge the gap between the real-time control domain and higher-level business systems and are commonly targeted in attacks aiming to disrupt OT or exfiltrate sensitive process information~\cite{kozak2023industroyer,sapalo2019vpnfilter,Salazar2024,mueller2016stuxnet,cosmicenergy}.

\subsection{Vulnerability Scanning as a Service}
\label{sec:scanners}
With the introduction of internet vulnerability scanners in the late 2000s, individuals have gained the ability to interrogate internet-connected devices at an unprecedented scale. 
Using specially crafted queries, these scanners systematically sample the global IP address space aiming to identify open ports and extract metadata about the services running on them. 
In exchange for a small-subscription fee, these platforms then provide searchable databases that allow users to query for specific protocols, devices and hosts discovered during scanning.
This enables users to assess the exposure of internet-connected systems at scale, with ease, and without their direct interaction with the target system. 
Popular internet-scanning providers include Shodan, Censys, Zoomeye, and FOFA, each offering different levels of scanning coverage and functionality~\cite{shodan,censys,Zoomeye,fofa}.

Although internet search engines advertise themselves as tools to help system owners assess their security posture, they also present significant risks. 
In practice, they may serve as a valuable tool for malicious actors to conduct broad reconnaissance for potential targets. 
Attackers could leverage these platforms to discover internet-exposed devices associated with a specific entity (such as an organization or geographic area) and quickly identify vulnerable services, providing an initial access point for further exploitation. 
This reconnaissance capability poses a particular risk to \glspl{ics}, as exposed endpoints (e.g., remote access services or misconfigured gateways) may allow attackers to pivot into an otherwise well-protected \gls{ics}, which is a commonly observed strategy~\cite{ncsc2022cyclopsblink,sapalo2019vpnfilter}.  
Additionally, opportunistic attackers can use these search engines to easily identify vulnerable devices at scale.
By querying for all instances of a specific service or device, a malicious actor may launch a widespread attack against many similar targets.

Due to these security implications, the legality of internet vulnerability scanners varies across jurisdictions. 
While some countries permit their use for research and security assessments, others impose restrictions or outright bans due to concerns over misuse. 
In addition to legal considerations, the research community continues to debate the ethical implications of these platforms and their methods~\cite{ndss-scanners-ethics}. 
Many scanners incorporate automated vulnerability detection, associating identified devices with known \glspl{cve} based on their advertised configurations. 
Again, while this feature can aid security professionals in identifying potential risks, it may also enable attackers seeking to exploit unpatched vulnerabilities at scale.
However, as discussed in Section~\ref{sec:methodology}, this functionality is only available for IT systems, likely due to the challenge in reliably scanning heterogeneous OT systems.

Recently, internet vulnerability scanners have enhanced their functionality through the introduction of machine learning algorithms capable of ingesting multimodal data from exposed services.
This represents a step change in the effectiveness of these scanners to discover some targets as they can now be identified through services exposing image or audio data.
Targets exposing an unsecured \gls{rdp}, \gls{gui}, webcam or microphone can now be categorized at scale, enabling malicious actors with a suite of new queries to discover potential targets.
This is particularly relevant for \gls{ot} security, as \glspl{hmi} and \gls{scada} systems commonly expose such services, or 
Previously, annotation of these targets would require manual effort or induce high false-positive rates, both of which prohibit efficacy of discovery at scale.

\subsection{The Influence of Honeypots on Internet Scanners}
Perhaps in response to the increasing prevalence of internet vulnerability scanners, network defenders and security researchers alike have explored the deployment of increasingly realistic honeypots. 
Honeypots are decoy systems designed to mimic real devices, attracting scanning activity and cyberattacks in order to collect intelligence on adversaries and subsequently improve intrusion detection. 
They serve as a fundamental tool for network defence, enabling researchers to study attacker behaviour, identify exploitation techniques, and enhance cybersecurity. 
The development of high-fidelity \gls{ics} honeypots, which closely replicate real industrial control devices, is an active research area~\cite{dodson,icspot,honeyplc,riotpot}. 

While internet scanners commonly implement automated techniques to detect and classify honeypots, these mechanisms are unlikely to be perfect, meaning that some honeypots may be misclassified as real devices and vice versa. 
This introduces uncertainty into the results of large-scale internet scanners, potentially affecting the accuracy of the assessments of the exposure of \gls{ics} devices presented in this work.

\section{Threat Model}
We consider a remote adversary targeting an \gls{ot} system, aiming to compromise the integrity and confidentiality of the physical process and its sensor data.
The adversary is assumed to have no physical access to the \gls{ot} environment but is capable of conducting reconnaissance over the internet to identify publicly exposed assets.
Such assets include target devices themselves (e.g., controllers, historians or \glspl{ews}) or services that enable remote access (e.g., \gls{rdp} servers, \gls{vpn} gateways or \glspl{hmi}). 
In the latter case, the exposed service itself is not the primary target, but may be exploited to enable lateral movement to the \gls{ot} network.

We consider several representative attack vectors.
First, an attacker may discover a remotely-accessible \gls{plc} supporting Modbus, an unauthenticated and unencrypted protocol.
The attacker may then forge or replay control messages to degrade the physical process performed by the \gls{plc} or extract sensor data.
Second, an attacker may discover a remotely-accessible \gls{gui} (e.g., \gls{scada} system or \gls{hmi}) that includes controls for a connected \gls{plc}, which itself is not exposed to the attacker. 
Through the \gls{gui}, the attacker may then view process information or issue commands to the connected \gls{plc}.
Finally, an adversary may exploit an internet-exposed \gls{it} device to gain initial access to a corporate \gls{it} network.
The attacker may then pivot through the \gls{it} network to discover an engineering workstation or \gls{scada} system with dual access to both IT and OT networks, subsequently issuing unauthorized commands to \glspl{plc}.
In each scenario, the attacker’s objectives may include disrupting or degrading the physical process, or extracting sensitive operational data.

\section{Methodology}
\label{sec:methodology}
To identify internet-exposed \gls{ot} devices, we utilise Shodan.
As discussed in Section~\ref{sec:scanners}, Shodan systematically scans a broad range of IP addresses and ports, sending crafted queries to detect accessible devices and services.
When a device responds, Shodan captures and stores the response as a "banner", a structured representation of the device's metadata.
Shodan was chosen for this study due to its global coverage, established reputation in both cybersecurity research and practical security assessments and low cost.
However, Shodan does present a limitation, unlike other internet scanning tools, it does not allow direct querying for many \gls{ot} services.
Instead, queries must be crafted based on common ports or identifying information within response banners.
As a result, devices that use non-standard ports or do not return uniquely identifiable metadata are challenging to measure precisely.
Further, Shodan (like other popular internet-device scanners) does not provide functionality for querying about specific \gls{ot} vulnerabilities.
Thus, the user must parse the devices' signatures to identify the manufacturer, model, and firmware versions and manually cross-reference these with vulnerability databases.
Such an analysis is performed in Section~\ref{sec:results}.

However, Shodan does provide supplementary tools that enhance analysis, including product-specific search filters for many \gls{ics} products and automated screenshot-capturing functionality, which is utilised in Section~\ref{sec:screenshots}.
We opt to focus on the most popular OT protocols to cover the largest attack surface.
Based on recent reports~\cite{koay2023machine,mirian2016internet}, we selected fifteen common OT protocols spanning different industries and devices.
A short description of each protocol is provided here, with its associated Shodan query included in Appendix~\ref{app:protocol-queries}.

\textbf{ModbusTCP:}
ModbusTCP is an open-source \gls{ics} protocol that allows a client to read and write memory addresses on a server, adapting the previous Modbus serial protocol to modern Ethernet networks.
Perhaps due to its simplicity, ModbusTCP has been adopted across a range of industries including energy, manufacturing, and building automation.


\textbf{EtherNet/IP:}
EtherNet/IP (Industrial Protocol) is a widely used industrial protocol based on the \gls{cip}, designed for automation and control applications.
EtherNet/IP implements two communication channels over Ethernet.
The first ensures timely communication of I/O data through the use of UDP, and the second communicates diagnostic data, where integrity is ensured through TCP.
The protocol is utilized by many vendors, including Rockwell Automation, Omron, and Schneider Electric, and is supported by a large ecosystem of interoperable devices.


\textbf{BACnet:}
BACnet is a communication protocol specifically designed for building automation and control systems. 
It facilitates interoperability between devices from different manufacturers in HVAC, lighting, and security systems. 
Due to its widespread adoption in smart buildings and commercial infrastructure, BACnet is a key protocol in facility management and automation.


\textbf{KNX:}
KNX is an open standard for home and building automation, enabling communication between smart devices such as lighting, heating, and security systems. 
It is widely used in residential, commercial, and industrial buildings, with support from major automation vendors. 
Its decentralised architecture and broad adoption make it an integral part of smart infrastructure.


\textbf{S7:}
S7 is a proprietary protocol developed by Siemens for communication between Siemens' \glspl{plc} and industrial automation devices. 
It is widely used in manufacturing, automotive, and critical infrastructure sectors.
Siemens' popularity in addition to the protocol's closed nature make it a frequent target in industrial cybersecurity research~\cite{alsabbagh2022remote,biham2019rogue7,alsabbagh2021stealth,hui2021vulnerability}.


\textbf{IEC 60870-5-104:}
IEC 60870-5-104 (IEC-104) is a standard protocol used for \gls{scada} systems in electrical power transmission and distribution.
It facilitates real-time communication between control centres, substations and \glspl{rtu}, making it common in grid management.
IEC-104 has been exploited on multiple occasions in cyberattacks aimed at energy distribution, namely by Industroyer in 2016 and Industroyer2 in 2022~\cite{kozak2023industroyer}.
Further malware samples exploiting IEC-104 have been discovered in public malware scanning databases~\cite{cosmicenergy}.


\textbf{CODESYS:}
CODESYS is an industrial-control software suite, including controller runtime software, controller programming software and remote access tools.
CODESYS aims to provide a hardware agnostic software solution, however it is often associated with Wago controllers.
The CODESYS network protocol, used to communicate between a CODESYS runtime and programming device, enables downloading/uploading of controller programs, monitoring, remote procedure calls and more. 
Its cross-vendor compatibility and software-driven nature make it a versatile but security-sensitive platform in industrial environments, that has been targeted for exploitation~\cite{pipedream}.


\textbf{Omron FINS}
Omron FINS is a proprietary protocol developed by Omron for communication between PLCs, HMIs, and industrial devices. 
It is predominantly used in manufacturing, logistics, and factory automation systems where Omron controllers are deployed. 


\textbf{Red Lion Controls:}
Red Lion Controls supports multiple industrial protocols, enabling interoperability between PLCs, HMIs, and SCADA systems. 
Its protocol stack facilitates communication in manufacturing, energy, and remote monitoring applications. 
Due to its multi-protocol nature, Red Lion devices often serve as protocol bridges in complex industrial networks.


\textbf{DNP3:}
\gls{dnp3} is a protocol designed for reliable communication in SCADA systems, particularly in electric power and water utilities. 
It is widely used in North America for monitoring and controlling remote industrial assets. 
Its design includes security extensions, but legacy implementations remain a concern in critical infrastructure.


\textbf{OPC UA:}
OPC UA (Open Platform Communications Unified Architecture) is a platform-independent communication protocol for industrial automation and \gls{iot} applications.
It enables data exchange between devices and systems across different vendors, making it a key in large and diverse industrial networks.
With built-in security features such as encryption and authentication, OPC UA addresses many challenges in modern industrial communication, however it has nonetheless been targeted by cyberattack in recent history~\cite{pipedream}.


\textbf{Unitronics PCOM:}
Unitronics PCOM is a proprietary protocol used for communication with Unitronics PLCs and HMIs. 
It is primarily used in manufacturing, process automation, and embedded control systems. 


\textbf{PCWorx:}
PCWorx is a protocol developed by Phoenix Contact for configuring and programming industrial automation systems. 
Common in factory automation and power distribution, PCWorx facilitates seamless integration with Phoenix Contact controllers. 


\textbf{NMEA:}
NMEA is a protocol standard used in marine electronics for communication between navigation devices, such as GPS receivers and shipboard sensors. 
It is widely adopted in the maritime industry for vessel monitoring, positioning, and environmental data collection. 
Due to its critical role in maritime safety, NMEA protocols are often scrutinized for security vulnerabilities~\cite{kessler2021can}.


\textbf{MELSEC-Q:}
MELSEC-Q is a proprietary protocol developed by Mitsubishi Electric for high-performance industrial automation systems. 
It is commonly found in manufacturing, robotics, and semiconductor industries, where Mitsubishi PLCs are deployed. 
With advanced control capabilities, MELSEC-Q is a key protocol in precision automation but requires careful security considerations in connected environments.


\section{Findings}
\label{sec:results}
\begin{figure}[!htbp]
\centering
\includegraphics[width=\linewidth]{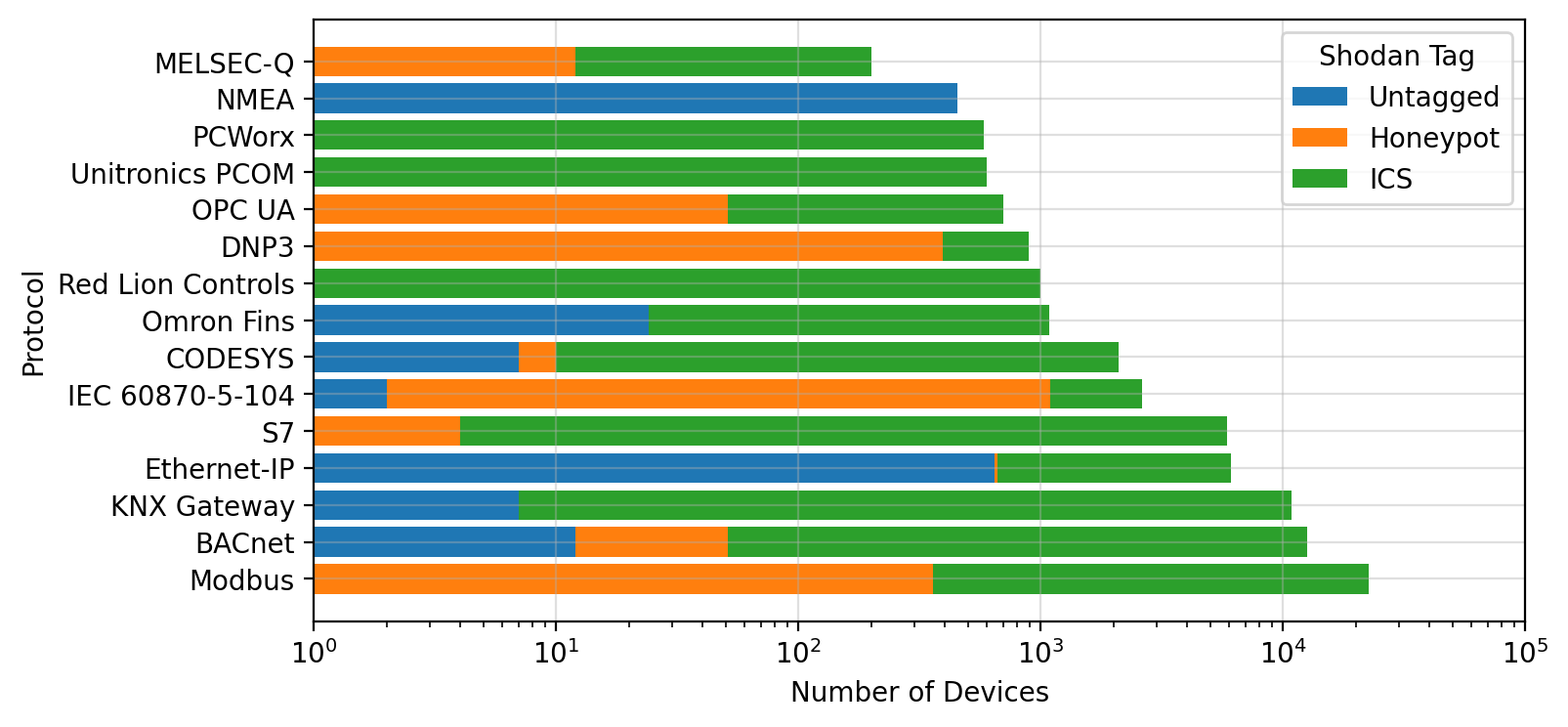}
\caption{Number of devices per protocol scan. Devices are categorised by their tag as assigned by Shodan's internal classification system.}
\label{fig:protocol-scan}
\end{figure}

Using the protocol queries from Section~\ref{sec:methodology}, we discover 68243 internet facing \gls{ot} devices globally, with counts per protocol shown in Figure~\ref{fig:protocol-scan}. 
Each protocol is annotated with the number of identified devices, in addition to honeypot and \gls{ics} \textit{tags}, a classification assigned by Shodan's proprietary algorithm.

Manual inspection of devices assigned the honeypot tag, indicated that this was applied to devices exhibiting signatures consistent with known honeypot implementations, such as the widely used Conpot ICS honeypot~\cite{conpot}. 
As the detection mechanism appears to rely on known signatures, this measure should be interpreted as a lower bound, as high-fidelity or recently developed honeypot implementations may evade detection. 
Broadly, the most common protocols scanned were not associated with a specific vendor, but rather represented widely adopted standards spanning multiple industries.

\begin{figure}[!htbp]
\centering
\includegraphics[width=0.99\linewidth]{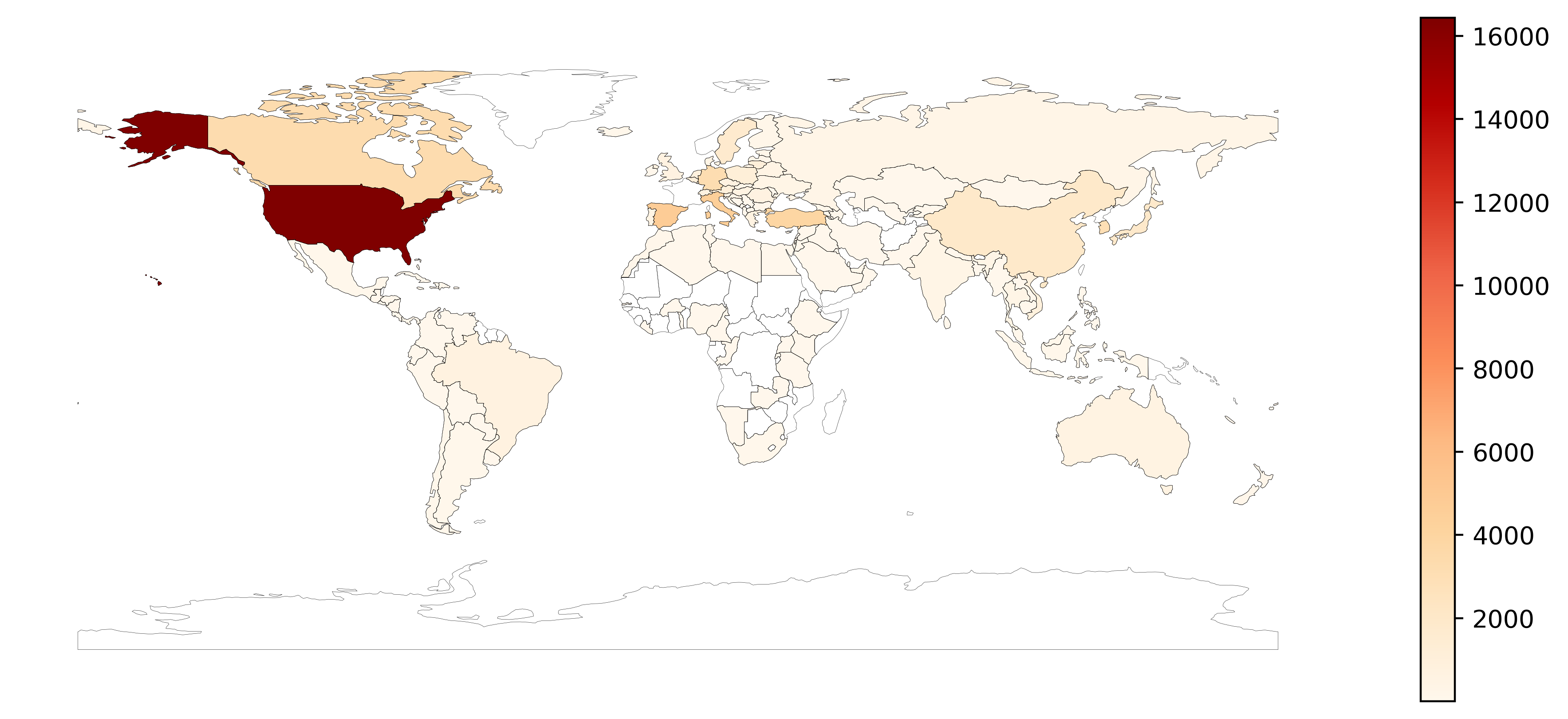}
\caption{Number of exposed \gls{ot} devices per country.}
\label{fig:protocol-map}
\end{figure}

The geographic distribution of devices discovered through scanning is illustrated in Figure~\ref{fig:protocol-map}. 
The map reveals that the largest concentrations of devices are found within North America and Europe, particularly in the United States, Canada, Turkey, Spain, Italy, and Germany.
However, it is important to note that the apparent geographic location of these devices may not always reflect their true physical locations.
Device owners can easily mask the location of their devices through the use of \glspl{vpn} or tunnelling through cloud service providers.

Following this broad analysis of exposed protocols, we examine devices exposing ModbusTCP, EtherNet/IP and S7 services in depth.
The selection of these protocols was based on their popularity, security implications and relevance to critical infrastructure.

\subsection{ModbusTCP Devices}
\label{sec:modbus-results}
The ModbusTCP protocol remains a widely used standard in industrial automation.
This is confirmed by our results in Figure~\ref{fig:protocol-scan}, which identified ModbusTCP as the most common internet-exposed \gls{ot} service.
This is concerning as ModbusTCP lacks many basic security features and thus may leave devices vulnerable to the following basic attacks:

\paragraph{Reconnaissance:} Due to its lack of encryption, ModbusTCP is vulnerable to man-in-the-middle attacks that may be used to capture sensitive data.
This could enable industrial espionage or IP theft by learning process parameters, or enable more complex attacks by discovering network endpoints and ModbusTCP-accessible memory addresses.
In 2018, this vulnerability was exploited by the VPNFilter malware~\cite{sapalo2019vpnfilter} which sniffed and exfiltrated ModbusTCP traffic.
\paragraph{Injection:} Due to its lack of authentication, ModbusTCP is vulnerable to injection attacks.
This may enable an attacker to inject malicious commands or spoof responses to compromise both ModbusTCP servers and clients.
While some protections are provided as ModbusTCP is encapsulated within a TCP connection, attackers may still perform man-in-the-middle injections, predict sequence numbers, or simply initiate their own session.
In 2024, this vulnerability was exploited by the FrostyGoop malware~\cite{frostygoop-palo-alto}, which injected ModbusTCP commands to disable heating within Ukrainian apartment buildings.
\paragraph{Denial of Service:} Attackers may overwhelm a ModbusTCP endpoint by injecting requests at a higher rate than the device can process.
While such an attack is easily detected and mitigated, it is similarly easily performed and may enable an attacker to temporarily disable a ModbusTCP endpoint.\\

\begin{figure}[!htbp]
\centering
\includegraphics[width=0.99\linewidth]{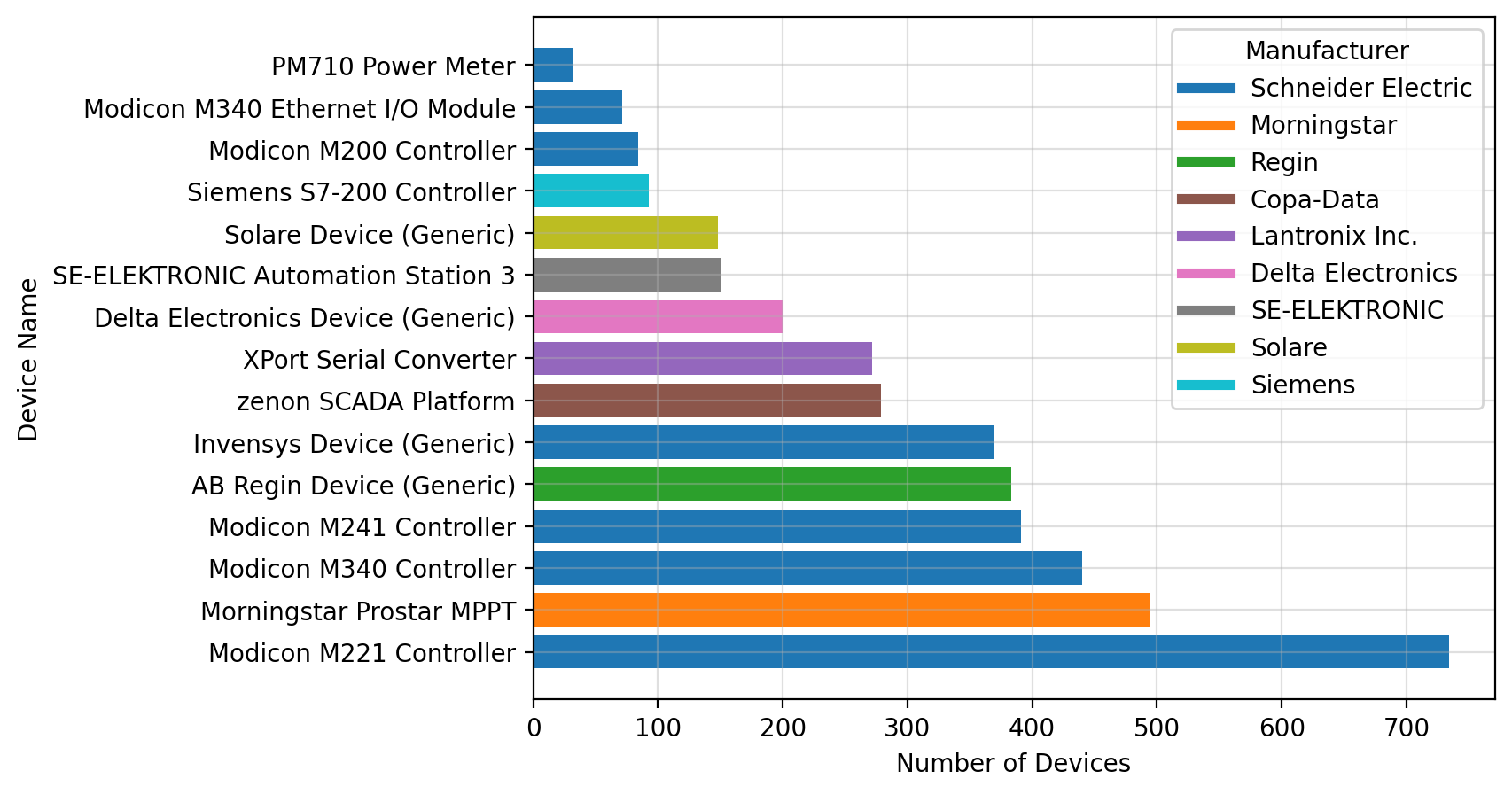}
\caption{15 most common devices reported by ModbusTCP ``Read Device Identification'' query.
706 unique devices were discovered from a total of 4967 responses.}
\label{fig:modbus-device-scan}
\end{figure}

Furthermore, ModbusTCP supports the ``Read Device Identification'' query, which enables devices to return information such as their make, model, and firmware version number.
While the majority of ModbusTCP devices discovered through our queries did not meaningfully respond to this self-identification request, approximately one fifth of devices returned identifying information, shown in Figure~\ref{fig:modbus-device-scan}.
Notably, nearly all devices omitted their firmware version, and many devices did not report specific model numbers, instead only returning a vendor name.
Devices within the latter category are denoted as ``generic''.

Among the devices that provided identifying information, Modicon \glspl{plc} were the most common, with the M221, M340, M241, and M200 models all ranking within the top fifteen.
Other frequently observed controllers included the Siemens S7-200 and SK-ELEKTRONIC's Automation Station 3.
Solar charge controllers were also frequently observed. Notably, the Morningstar ProStar \gls{mppt} and Solare (Generic) devices were among the most common, indicating a trend of ModbusTCP adoption in renewable energy applications.
The exposure of such devices to the internet presents risks, as attackers could possibly manipulate these controllers en masse to disrupt power grid infrastructure.

Devices that report detailed information about their manufacturer and model may facilitate reconnaissance for potential attackers, presenting an unnecessary risk.
This may enable large-scale attacks, as device-specific vulnerabilities can easily be scaled to all compatible devices.
Such an attack could utilise an alternative service rather than ModbusTCP, such as a vendor-specific controller service, which may now be enabled due to ModbusTCP reporting detailed device information.
Device manufacturers might consider reporting only coarse device information to help protect devices, as adversaries must then develop fingerprints that infer device models based on network behaviour, response patterns, or other metadata.
In either case, once a device model is identified, an attacker can leverage vulnerability databases to apply known exploits.
For example, the Modicon M221, the most commonly observed ModbusTCP device, has multiple CVEs that could be exploited if the device is unpatched \cite{modicon221-cve}.
Beyond known vulnerabilities, attackers can utilise manufacturer documentation to gain insight into device behaviour and potential weaknesses.
For instance, the ProStar MPPT solar charge controller, the second most frequently observed device, has detailed ModbusTCP communication documentation available online \cite{prostar-documentation}, which could be used to craft targeted attacks without requiring reverse engineering.

\subsection{EtherNet/IP Devices}
\label{sec:ethernet-results}
\begin{figure}[!htbp]
\centering
\includegraphics[width=1.0\linewidth]{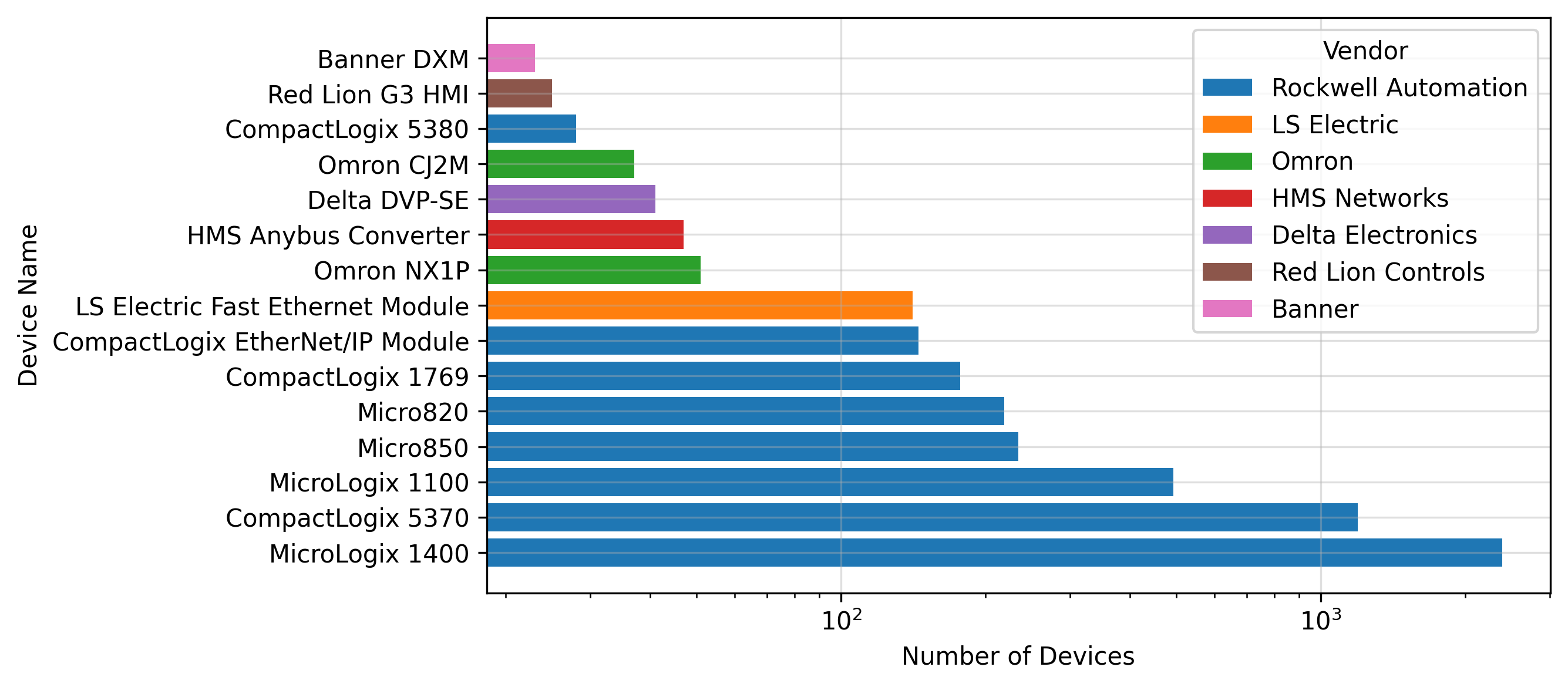}
\caption{The 15 most common device manufacturers and models reported by EtherNet/IP scanning.}
\label{fig:ethernet-ip-vendors}
\end{figure}

EtherNet/IP is a widely used industrial protocol that shares many of the security vulnerabilities of ModbusTCP, as it lacks encryption and authentication in its default configuration.
However, EtherNet/IP offers a security extension, \gls{cip} Security, which provides authentication and encryption through \gls{tls} when enabled.
Banner information provided by Shodan did not reveal whether EtherNet/IP devices had such security features enabled.
Unlike scanning results for ModbusTCP, nearly all EtherNet/IP devices reported their make and model numbers, and many included firmware versions.
The most frequently observed device models are shown in Figure~\ref{fig:ethernet-ip-vendors}.


\begin{figure}[!htbp]
\centering
\includegraphics[width=0.8\linewidth]{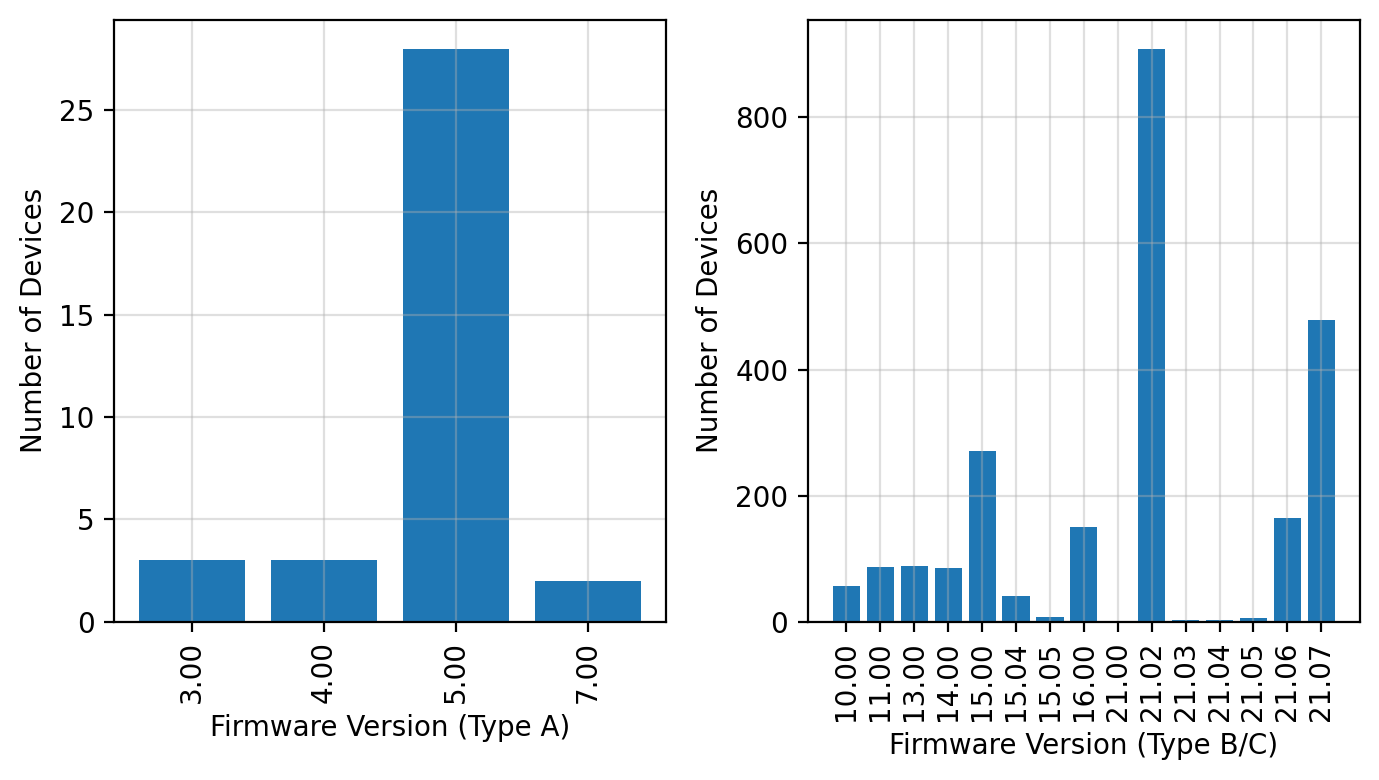}
\caption{Firmware versions of MicroLogix 1400 controllers.}
\label{fig:micrologix-firmware-versions}
\end{figure}

Rockwell Automation controllers were the most common devices, specifically from the MicroLogix, CompactLogix, and Micro series.
In particular, the MicroLogix 1400 controller was the most common, and additionally reported firmware version numbers, as shown in Figure~\ref{fig:micrologix-firmware-versions}, enabling reference with public vulnerability databases.
The MicroLogix 1400 comes in A, B and C types, denoting the generation of device releases.
All Type A controllers discovered by our scans appear vulnerable to a remote code execution, flaw affecting firmware versions $\leq$ 7.0~\cite{cisaRockwellAutomation}.
While the contemporary Type B and C controllers incorporate improved security measures, remote code execution vulnerabilities persist in firmware versions 21.02 and older, impacting approximately 1200 of the discovered devices~\cite{micrologix1400-cve-21.2}.
This vulnerability was disclosed in January 2018, highlighting that the majority of affected device owners have failed to patch this critical security vulnerability more than seven years following its public disclosure.

Type B/C devices running firmware versions $\leq$ 21.06 remain susceptible to an additional exploit discovered in July 2021, which allows an attacker to read and write controller registers using a crafted ModbusTCP query~\cite{micrologix1400-cve}.
This demonstrates how insecure industrial protocols may be used in conjunction, where reconnaissance may be performed with one protocol enabling exploitation by another.
Despite the availability of a patch in July 2021, nearly 200 identified devices continue to run unpatched firmware, again demonstrating the reluctance for OT device owners to perform security maintenance. 
Further, at the time of writing, Rockwell Automation still provides customers with the option to download the vulnerable 21.06 firmware and, unlike its advisory for the Type A controller vulnerabilities, does not appear to have issued a security warning regarding this flaw.
This raises general concerns about vendor responsibility in mitigating known risks and ensuring that asset owners are adequately informed of critical security threats associated with their devices.

\subsection{S7 Devices}
\label{sec:s7}
\begin{figure}[!htbp]
\centering
\includegraphics[width=1.0\linewidth]{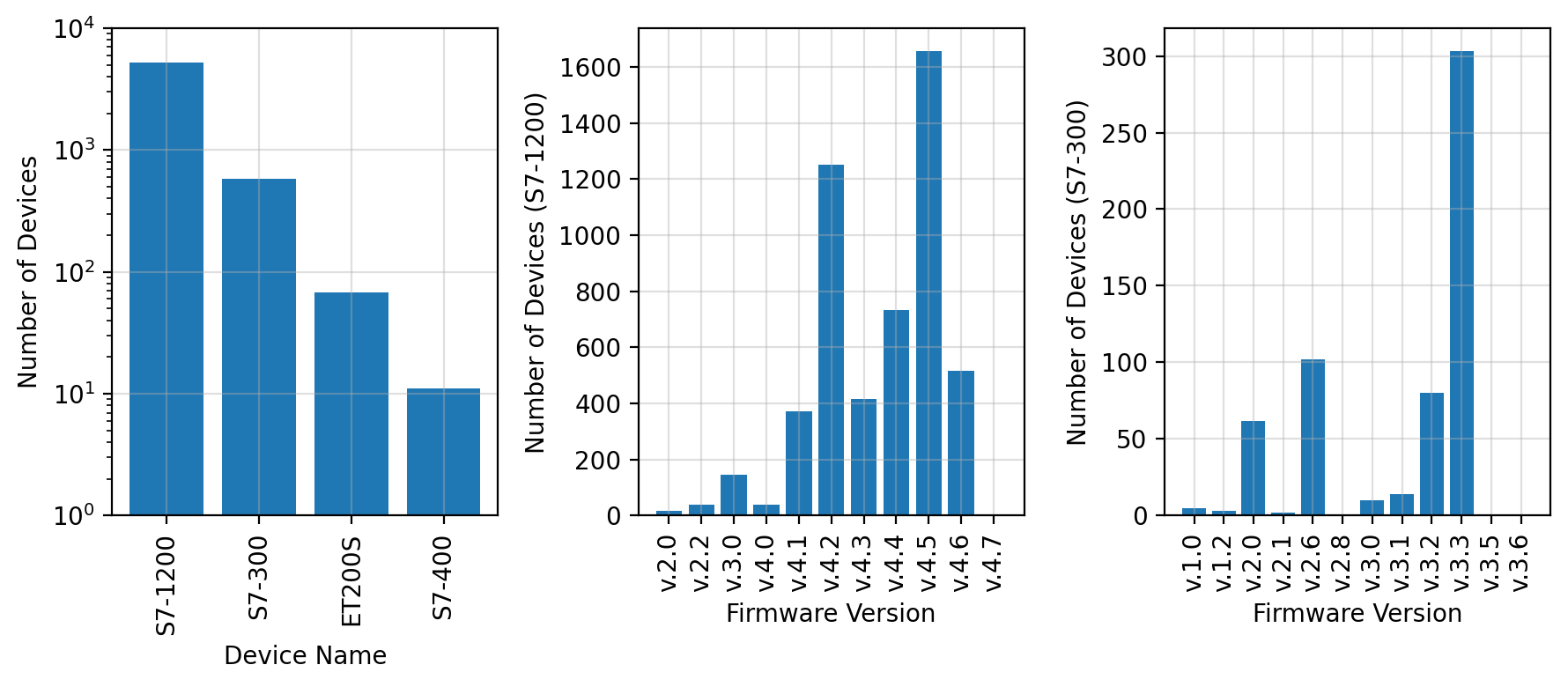}
\caption{S7 devices discovered and firmware versions of S7-1200 and S7-300 \glspl{plc}.}
\label{fig:s7-firmware}
\end{figure}

S7 is a proprietary network protocol used by Siemens' Step7 programming software to facilitate communication with Siemens \glspl{plc}.  
Through this protocol, we examine a critical attack vector within \glspl{ics}, where \gls{ews} software may be exploited by an attacker.
This is particularly concerning as \gls{ews} software often implements specialised functionality that allows increased control and monitoring of devices through proprietary protocols.  
In many cases, this software also enables the reprogramming of controllers, making it a valuable target for exploitation.
While this analysis focuses on S7, similar architectures exist across other vendors, exposing comparable security risks (eg. CODESYS, Omron FINS, Unitronics PCOM, MELSEC-Q).
The models of S7-compatible devices identified in our scanning results are shown in Figure~\ref{fig:s7-firmware}.  

One of the primary security concerns with S7 is its lack of encryption and authentication, making it susceptible to the vulnerabilities discussed in Section~\ref{sec:modbus-results}.  
Additionally, as a proprietary protocol, S7 poses challenges for detection and mitigation, as network administrators do not have access to detailed protocol documentation to decode traffic effectively. 
In response to these vulnerabilities, Siemens introduced S7CommPlus in 2014, incorporating encryption and authentication through a firmware update.  
This security enhancement is implemented in S7-1500 and S7-1200 PLCs running firmware versions $\geq$ 4.0.  
However, this update was not compatible with older S7-300 controllers, which remain vulnerable to sniffing and injection attacks.
The firmware versions of S7-1200 and S7-300 controllers, the two most common S7 controllers discovered, are shown in Figure~\ref{fig:s7-firmware}.

Despite these security improvements, vulnerabilities persist in later firmware versions of S7 devices, albeit with greater exploitation difficulty.  
For example, version 4.1 of the S7-1200 firmware contains flaws that allow attackers to bypass authentication mechanisms and execute arbitrary commands~\cite{blackhat2017s7commplus}.  
Additionally, S7-1200 firmware versions $<$ 4.5 contain weakly protected credentials, making them susceptible to credential harvesting attacks~\cite{cisa2022s7commplus}.
While many asset owners have applied firmware updates to mitigate these risks, approximately half of the devices discovered were unpatched.  
This indicates that although Siemens devices tend to be more regularly updated as compared to other industrial controllers, they still exhibit widespread security gaps.  

\subsection{Automated Discovery through Images}
\label{sec:screenshots}
\begin{figure}[!htbp]
\includegraphics[width=.24\linewidth]{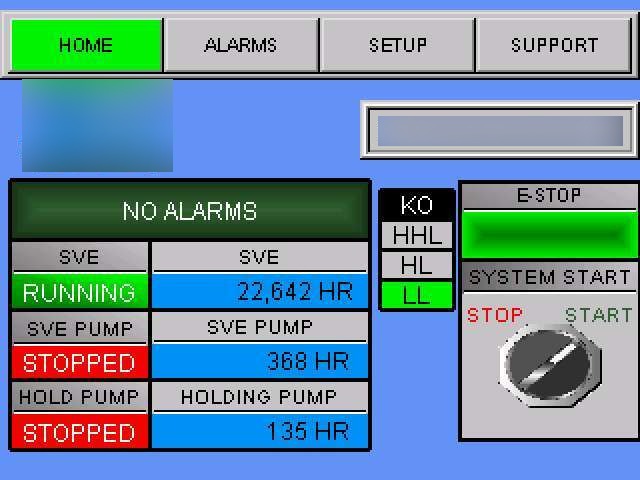}
\includegraphics[width=.24\linewidth]{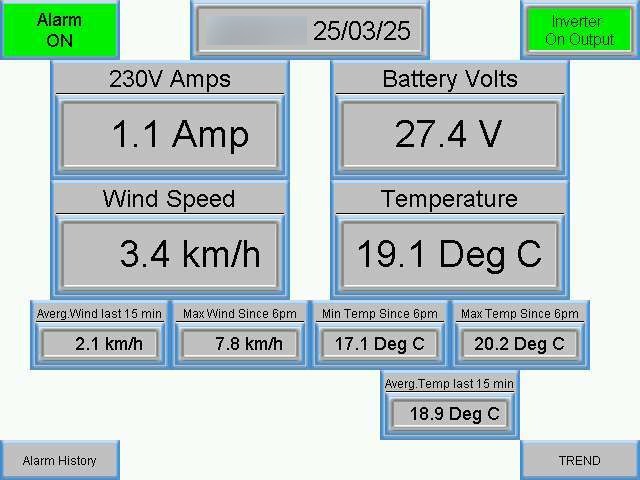}
\includegraphics[width=.24\linewidth]{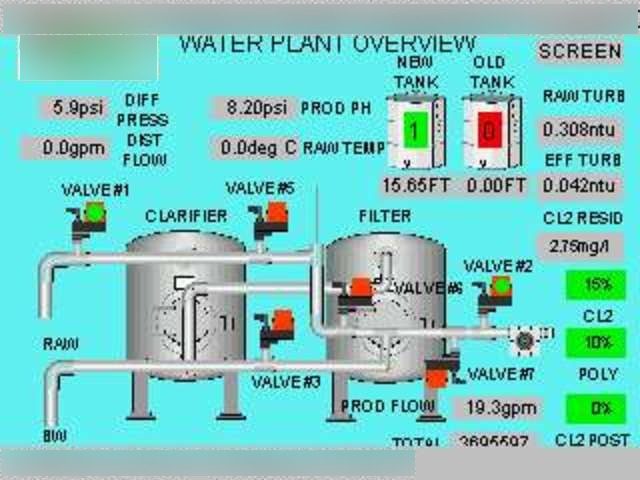}
\includegraphics[width=.24\linewidth]{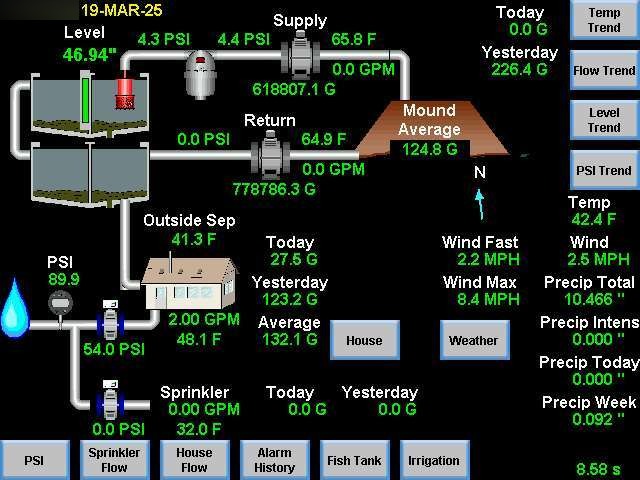}
\caption{HMIs identified by Shodan's ICS image classifier.}
\label{fig:ics-classifier}
\end{figure}

\begin{figure}[!htbp]
\centering
\includegraphics[width=0.24\linewidth]{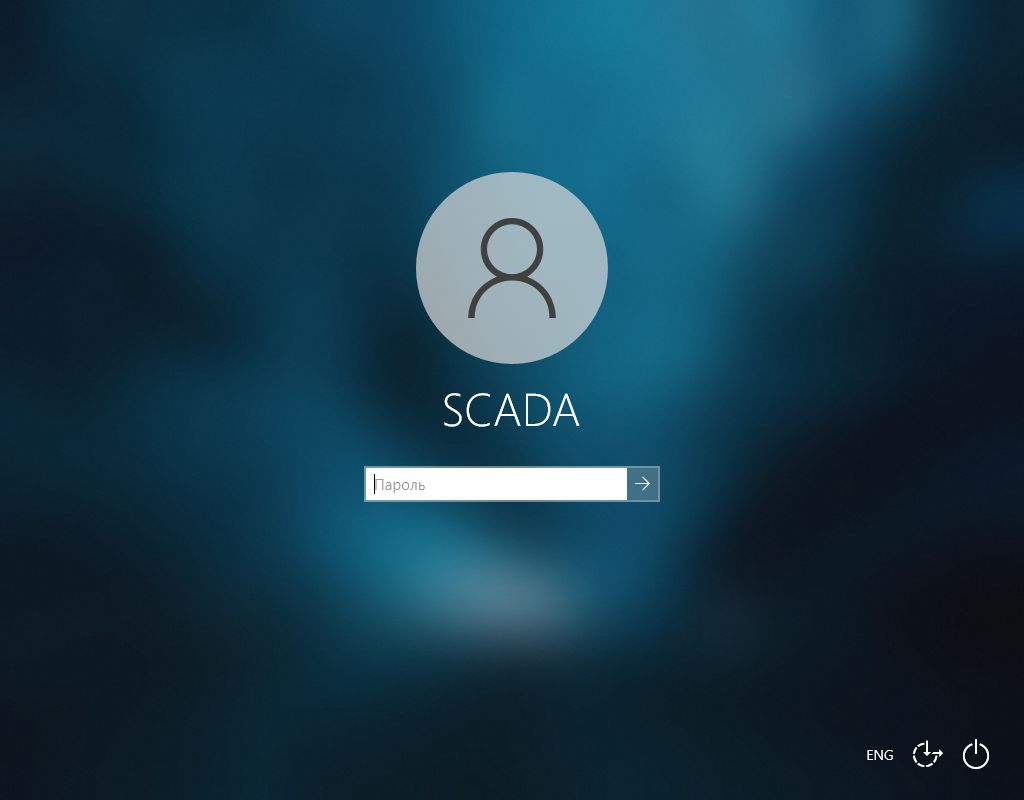}
\includegraphics[width=0.24\linewidth]{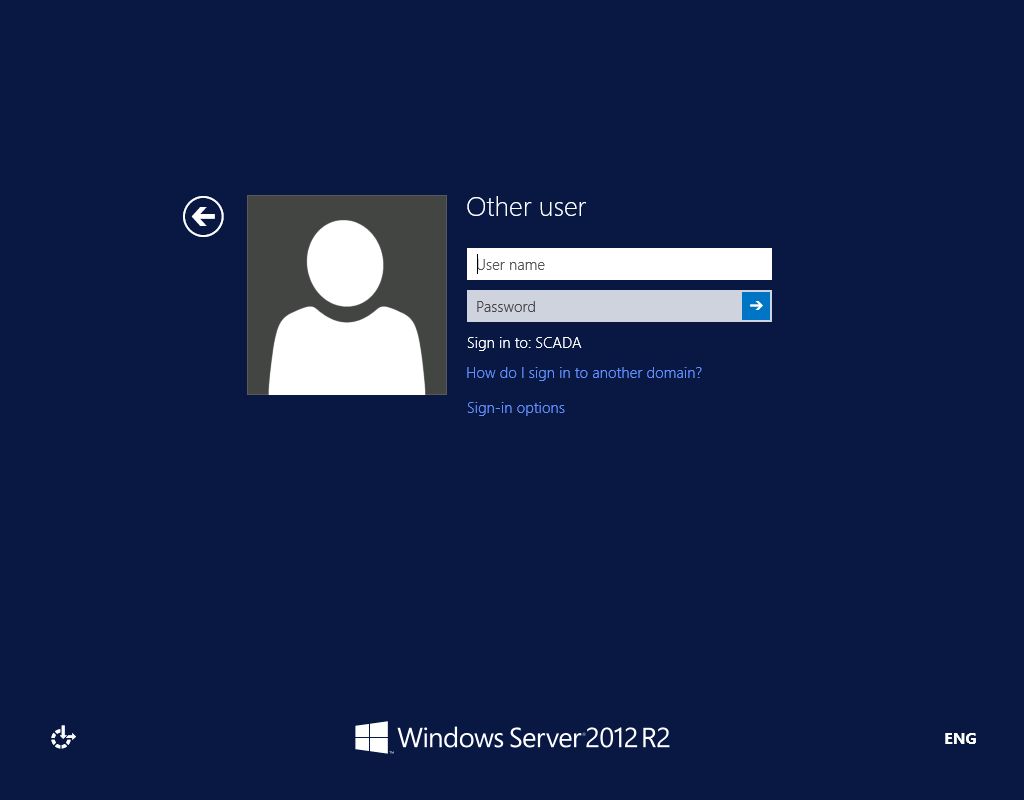}
\includegraphics[width=0.24\linewidth]{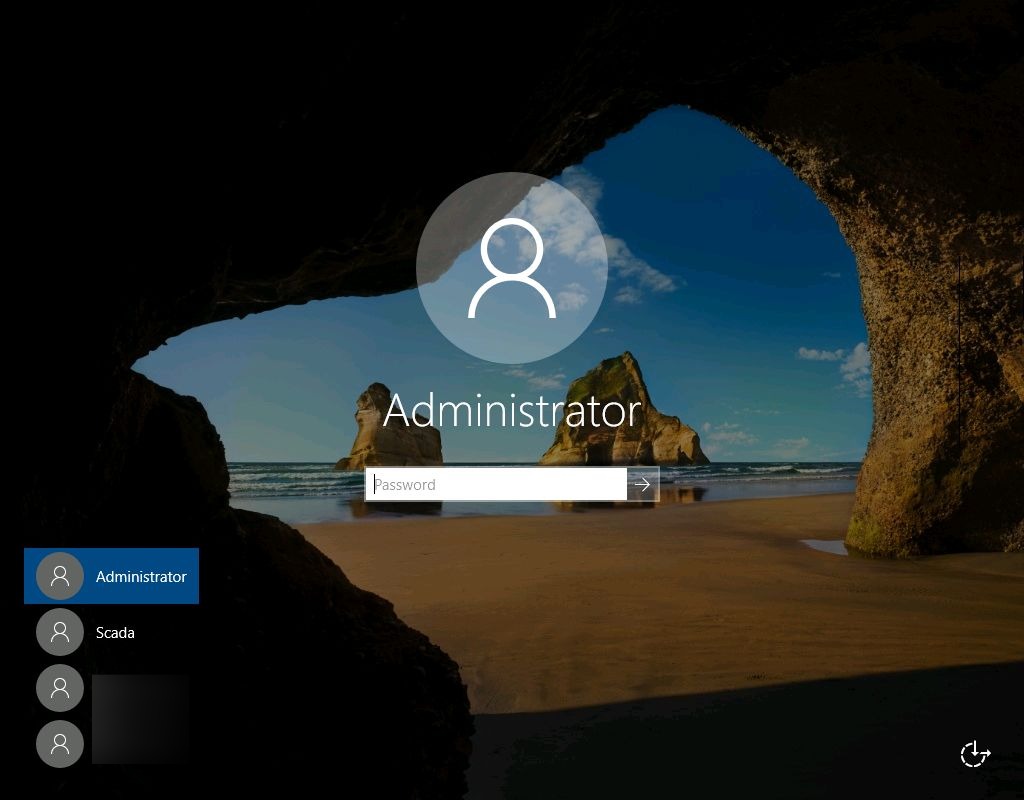}
\includegraphics[width=0.24\linewidth]{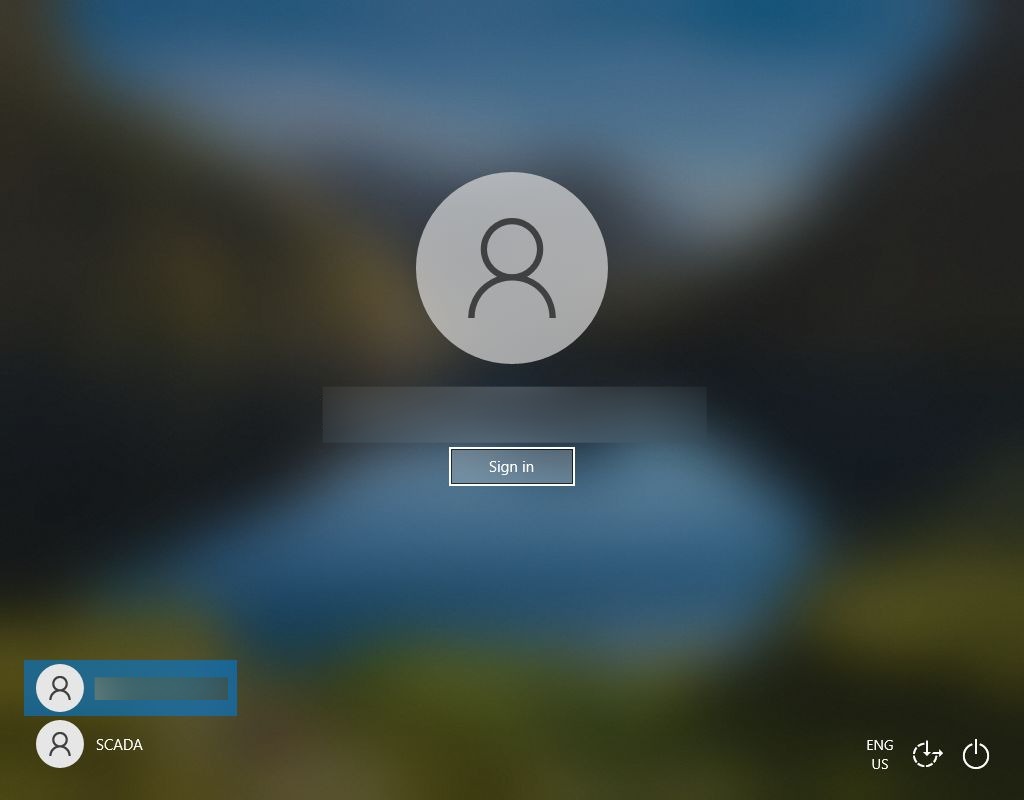}
\caption{SCADA systems identified through Shodan's OCR functionality.}
\label{fig:scada}
\end{figure}

\begin{figure}[!htbp]
\centering
\includegraphics[width=0.24\linewidth]{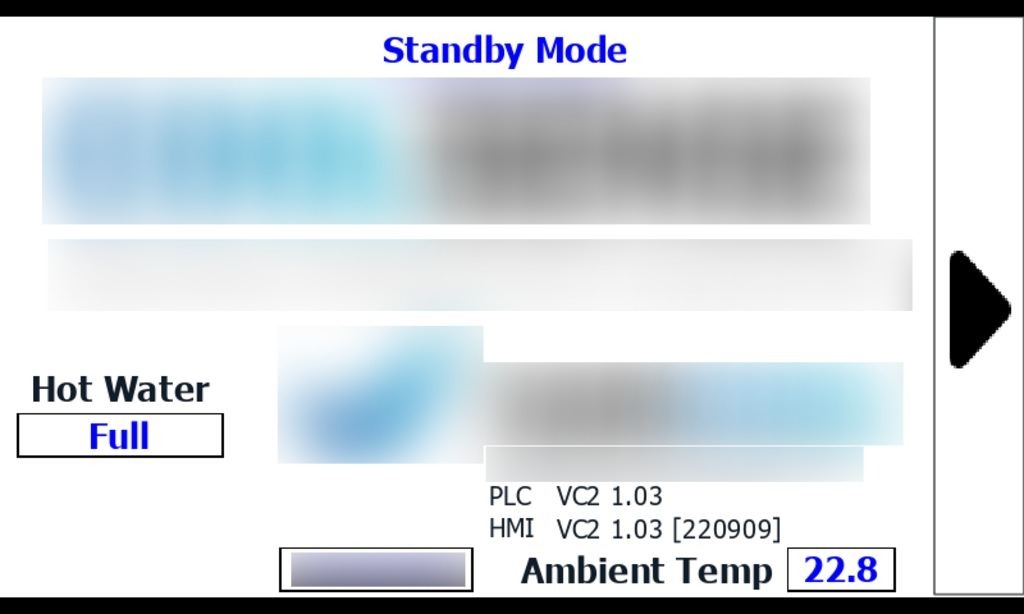}
\includegraphics[width=0.24\linewidth]{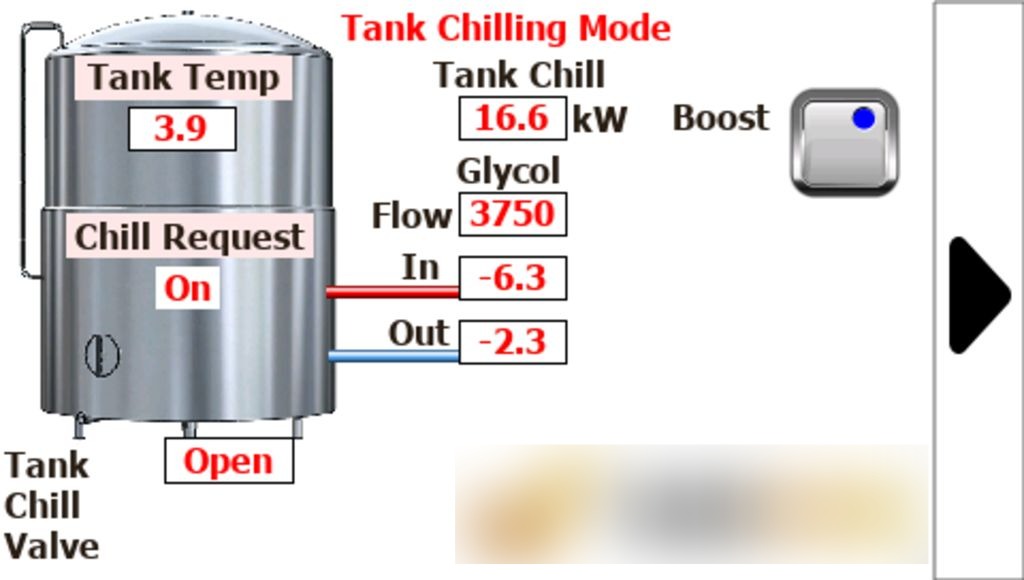}
\includegraphics[width=0.24\linewidth]{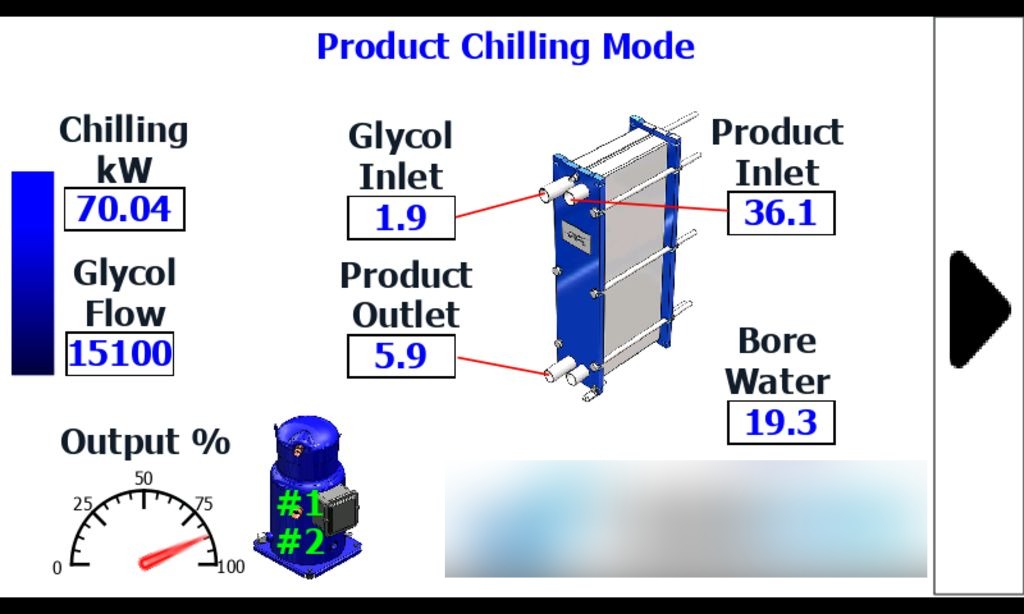}
\includegraphics[width=0.24\linewidth]{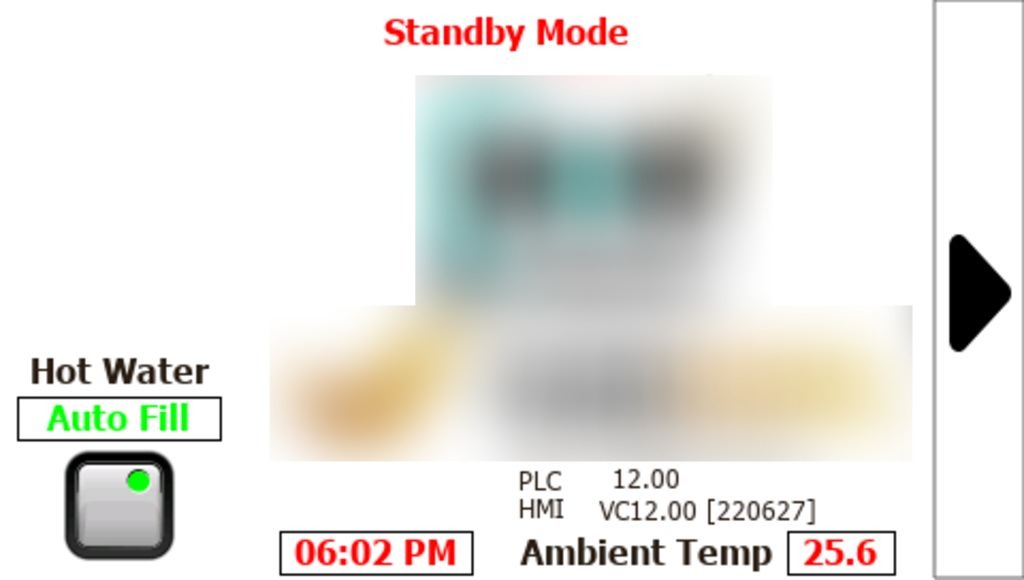}
\caption{HMIs identified through Shodan's OCR functionality by querying for ICS-specific keywords (``HMI'', ``PLC'', ``Temp'' and ``Flow'').}
\label{fig:hmis}
\end{figure}

Shodan includes functionality to enable querying over screenshots, a capability that is particularly relevant in \gls{ot} environments, graphical-only services (such as \gls{rdp}) are commonly implemented.
To enhance users' searches, Shodan provides an image classifier that automatically annotates images collected from such services.
One such annotation is \textit{ICS}, which can discover unsecured \gls{ot} devices such as \glspl{hmi}, as shown in Figure~\ref{fig:ics-classifier}.
Previously, discerning \gls{ot} devices from \gls{it} devices through these graphical services would be challenging, allowing \gls{ot} some safety through obfuscation.
In addition to image classification, Shodan provides \gls{ocr} to enable querying over text present within images.
A sample of results collected using this functionality is shown in Figure~\ref{fig:scada}, with a similar query for additional \gls{ics} terms presented in Figure~\ref{fig:hmis}.
The Shodan query syntax to perform \gls{ics} image classification and \gls{ocr} search is provided in Appendix~\ref{app:image-queries}.

\section{Discussion and Mitigations}
\label{sec:discussion}
With the introduction of remote workforces, IoT devices, and the adoption of Industry 4.0, \gls{ot} asset owners are increasingly incentivised to connect their devices to the internet to improve efficiency and reduce costs.
For instance, a survey performed by Fortinet in 2024 found 54\% of businesses have implemented remote \gls{ot} access, increasing from 40\% in 2022~\cite{Fortinet2024}.
In response to this shift, device vendors are gradually implementing secure-by-design principles and basic security enhancements to traditionally insecure OT protocols.  
Despite these advancements, the findings presented in Section~\ref{sec:results} indicate that legacy devices and insecure protocols still dominate.
Furthermore, asset owners appear reluctant to apply security patches, even when critical vulnerabilities have been publicly disclosed for several years.
 
Asset owners may be unaware of their exposure due to what appears to be a lack of security notification from device manufacturers, as demonstrated in Section~\ref{sec:ethernet-results}.
This may contribute to the varied security posture observed across different device vendors, for instance Rockwell Automation and Siemens users.
Additionally, the operational downtime required for patching may deter asset owners from prioritising security updates due to financial constraints. 
A further concern is system-owner's distrust of patching processes due to concerns that updates could disrupt existing operation.

Despite these challenges, the most straightforward and potentially effective security measure remains restricting public access to OT assets.
Where feasible, OT environments should consider limiting direct internet exposure and adhere to the Purdue Model, ensuring proper network segmentation to minimise their attack surface.
For systems where such segmentation is impractical, asset owners should consider implementing tunnelling mechanisms to compensate for the lack of encryption and authentication in many OT protocols.  
By implementing network isolation and secure communication channels, OT asset owners can significantly reduce the risk of unauthorized access and cyber threats to critical infrastructure.



\section{Conclusion}
\gls{ot} security is currently in a transitory period, with manufacturers beginning to implement secure-by-design devices that significantly enhance the security posture of industrial environments.  
These advancements represent a critical step toward mitigating long-standing vulnerabilities in OT infrastructure.  
However, the extended lifespan of OT devices results in a reliance of many asset owners on insecure legacy systems for the near future.  
As a result, outdated devices and protocols remain widely deployed, exposing critical infrastructure to critical cyber threats.  

Compounding this challenge, malicious actors now have unprecedented access to reconnaissance tools that allow them to discover and scale attacks across similar internet-connected devices.  
Through our scans of 15 common OT protocols, we demonstrate that many asset owners remain at significant risk of compromise.  
Despite the increasing availability of security enhancements, a substantial number of devices continue to expose insecure services.
Additionally, our findings reveal a widespread reluctance to implement security patches, leaving known vulnerabilities unaddressed for years following their disclosure.  


\bibliographystyle{splncs04}
\bibliography{bibliography}

\clearpage
\appendix
\section{\gls{ot} Protocol Shodan Queries}
\label{app:protocol-queries}
The Shodan queries for device enumeration across the 15 protocols described in Sec~\ref{sec:methodology} are shown in Table~\ref{tab:protocols}.
Quotation marks within queries denote the term is used to search banner information, whereas unquoted terms use Shodan's internal device detection functionality.

\begin{table}
\centering
\caption{Shodan queries for \gls{ot} network protocols.}\label{tab:protocols}
\begin{tabular}{|r|l|l|}
\hline
\textbf{Index} & \textbf{Protocol} & \textbf{Shodan Query} \\
\hline
1 & ModbusTCP & \queryfmt{port:502 "Unit ID" -http -ssl} \\
2 & EtherNet/IP & \queryfmt{port:44818,2222 "product name"} \\
3 & BACnet & \queryfmt{port:47808 "Instance"} \\
4 & KNC & \queryfmt{KNX Gateway} \\
5 & S7 & \queryfmt{port:102 "Basic Hardware"} \\
6 & IEC 60870-5-104 & \queryfmt{port:2404 asdu address} \\
7 & CODESYS & \queryfmt{CODESYS:} \\
8 & Omron FINS & \queryfmt{port:9600 response code} \\
9 & Red Lion Controls & \queryfmt{port:789 "Red Lion Controls"} \\
10 & DNP3 & \queryfmt{port:20000 "source address"} \\
11 & OPC UA & \queryfmt{port:4840 DisplayName} \\
12 & Unitronics PCOM & \queryfmt{"Unitronics PCOM"} \\
13 & PCWorx & \queryfmt{PCWorx} \\
14 & NMEA & \queryfmt{Gpgga Gprmc} \\
15 & MELSEC-Q & \queryfmt{port:5006,5007 product:mitsubishi} \\
\hline
\end{tabular}
\end{table}

\section{Shodan Queries for Image Processing}
\label{app:image-queries}
The Shodan queries for discovering ICS devices with image processing, as described in Sec~\ref{sec:screenshots} are shown in Table~\ref{tab:ics_queries}.
\queryfmt{<search\_terms>} denotes a comma separated list of keywords to search \gls{ocr} results.

\begin{table}
\caption{Shodan query techniques for ICS identification using images.}\label{tab:ics_queries}
\centering
\begin{tabular}{|r|l|l|}
\hline
\textbf{Index} & \textbf{Description} & \textbf{Query} \\
\hline
1 & ICS image classifier & \queryfmt{screenshot.label:ics} \\
2 & OCR content search & \queryfmt{has\_screenshot:true <search\_terms>} \\
\hline
\end{tabular}
\end{table}

\end{document}